\documentclass[preprintnumbers, twocolumn,showkeys,superscriptaddress,showpacs,
nofootinbib,prd,notitlepage]
{revtex4-1}
\usepackage{bbm,pifont}
\usepackage{amssymb}
\usepackage{amsmath}
\usepackage{physics}
\usepackage{enumitem}
\usepackage{amsfonts}
\usepackage{epsfig}
\usepackage{graphicx}
\usepackage{tabularx}
\usepackage{color}
\usepackage{slashed}
\usepackage[colorlinks=true, linkcolor=blue, citecolor=blue, urlcolor=blue]{hyperref}   

\begin{document}
\title{Radiation of relativistic electrons created in tunnel ionization of atomic gases\\
by laser beams of extreme intensity}

\author{N.V. Makarenko}
    \email[Correspondence email address: maknik1111@yandex.ru]{}
    \affiliation{National Research Nuclear University MEPhI, Kashirskoe shosse 31, 115409, Moscow, Russia}

\author{S.V. Popruzhenko}
    \email[Correspondence email address: ]{sergey.popruzhenko@gmail.com}
    \affiliation{National Research Nuclear University MEPhI, Kashirskoe shosse 31, 115409, Moscow, Russia}
    \affiliation{Institute of Applied Physics  RAS, Ul'yanova 46, 603950, Nizhny Novgorod, Russia}

\date{\today} 

\begin{abstract}
We consider tunnel ionization of atomic argon in a femtosecond laser pulse of intensity $10^{21}-10^{22}{\rm W/cm}^2$ aiming to investigate the relativistic dynamics and radiation of photoelectrons released from their parent ions inside the laser focus.
Radiation of such electrons accelerated along the laser field propagation direction appears to have moderate power but can be considerably enhanced by a collision with a relatively weak counter-propagating laser pulse. 
Using the theory of laser-induced tunneling in atoms and ions and that of nonlinear Thomson scattering, we demonstrate that angular distributions and spectra of emitted photons can serve as a probe of the peak intensity in the focus.
The angular distribution of emitted radiation in the plane of polarization and propagation of the ionizing laser beam appears narrow and peaked around an intensity-dependent angle, making this ionization setup a source of collimated XUV radiation.

\end{abstract}

\keywords{intense laser radiation, tunnel ionization, radiation, nonlinear Thomson scattering}

\maketitle

\section{Introduction}\label{sec:intro}

The recent development of the new generation of femtosecond laser sources with peak power in the interval 1-10 petawatts (PW) has opened a way to systematic laboratory studies of laser-matter interactions at intensities $10^{22}{\rm W/cm}^2$ and higher.
Presently, several laser facilities \cite{apollon,corels,ELI-BL,ELI-NP,SULF,apollon_25} are already capable of delivering pulses of 3-5PW power.
The peak power of 10-15~PW at a pulse duration of 20-30~fs looks realistic for the near future.
Projects of multi-beam sources \cite{sel_21,XCELS} open a potential possibility to reach the sub-exawatt level of power.
For a review of the current state of the art of the high-power laser sources and of the future projects, see recent publications \cite{gonoskov_rmp22,fedotov_pr23}.

These advances in laser technologies are expected to make the intensity domain $10^{22}-10^{24}{\rm W/cm}^2$ accessible for regular experiments soon.
At such extreme intensities, the nonperturbative regime of light-matter interaction becomes dominant, and the electromagnetic field strength and its distribution inside the focal volume are the key characteristics that determine probabilities of nonlinear phenomena of classical and quantum electrodynamics.
Therefore, reliable and versatile diagnostics of the laser focus are in high demand for forthcoming experiments on the interaction of extremely strong electromagnetic fields with charged particles, atoms, molecules, and plasmas.

Several methods to measure the peak intensity of laser radiation and its distribution in the focus have been recently proposed and discussed in the context of multi-PW installations, see e.g. \cite{he_oe19,mackenroth_njp19,yandow_pra19,vais_apb19, poprz_pra19,vais_njp20,popruzhenko_usp23} and references therein.
The one based on the observation of multiple tunnel ionization of heavy atoms \cite{walker_pra01,ueda_pra03,link_instr06,poprz_pra19,ciappina_lpl2020,mironov_pre25} suggests a way to calibrate the peak intensity with precision on the level of $\approx 20-50\%$ by observing the highest charge states of a given atomic specie.
Numerical calculations made for argon, krypton, and xenon \cite{poprz_pra19,ciappina_lpl2020,mironov_pre25} demonstrated its potential accessibility. 
Although such relatively low accuracy withdraws the method from the list of high-precision protocols of measurements customary to nonlinear optics and spectroscopy, it remains sufficient for rough estimates of the conditions achieved in the focal spot of an extremely intense laser beam.
The scheme of intensity calibration through the observation of multiple ionization of heavy atoms is now under consideration at several multi-PW laser installations, and the process of tunnel ionization is included into particle-in-cell codes used for simulations of ultrafast plasma dynamics in the field of intense laser radiation, see the recent publication \cite{mironov_pre25} for further details and references.

Except for possible applications in diagnostics, ionization of deeply bound states in laser fields of ultrahigh intensity can be used as a source of seed electrons for laser–plasma accelerators \cite{pak_prl2010,mcguffey_prl2010,clayton_prl10} and for initiation of quantum-electrodynamical cascades of elementary particles \cite{tamburini_sr17,artemenko_pra2017}.
The post-ionization dynamics of photoelectrons and the spectra of emitted high-energy photons enter as input data in simulations of these multi-particle phenomena.

In this paper, we address an effect directly connected with ionization in the focus of extremely powerful laser beams: electrons liberated from ions through tunneling will  accelerate in the same electromagnetic field, which caused the ionization, and emit electromagnetic waves predominantly in the direction of the laser pulse propagation. 
For a given atomic species, spectra and angular distributions of this radiation will depend on the peak laser intensity, on its distribution within the focus, and on the laser pulse duration.
This therefore provides an additional opportunity for probing the electromagnetic field structure in the focal volume.
{ It is essential that this ionization-induced radiation can be analyzed in the same experiment designed to study multiple ionization and the subsequent laser acceleration of photoelectrons.}

Here we analyze this ionization-induced radiation in view of the aforementioned perspective.
We show that the co-propagating motion of photoelectrons, although accelerating them up to ultra-relativistic energies, suppresses radiation considerably so that only a few high-energy photons will be emitted per single atom.
A relatively weak counter-propagating probe pulse is needed to enhance the emitted energy. 
The spectral and angular structure of this signal and its sensitivity to the time delay between the two pulses will also be addressed in our calculations.

The paper is organized as follows.
In the next section, we formulate the statement of the problem, introduce basic equations and discuss parameters of the laser pulses, of the atomic target and approximations used in our calculations.
Section III presents analysis of the radiation spectra with and without the probe counter-propagating pulse.
The last section delivers conclusions and an outlook.

\section{Statement of the problem and basic equations\label{sec:Statement}}

We consider radiation of electromagnetic waves by fast electrons born in the laser focus through ionization of argon ions at low pressures (see the sketch in Fig.1).
{ Argon is chosen for the following reasons.
Firstly, noble gases are commonly used in strong-field ionization experiments, see e.g. \cite{walker_pra01,ueda_pra03,link_instr06}, for they are easy to store and to operate with in a chamber.
Secondly, ionization potentials of the inner shells of these atoms are well adjusted to probe laser intensities in different intervals of the wide $10^{21}-10^{24}{\rm W/cm}^2$ domain.
In particular, the $1s$ shell of argon is depleted due to field ionization at intensity $\approx 4\cdot 10^{21}{\rm W/cm}^2$ \cite{poprz_pra19}.
The concentration of atoms in the interaction area is assumed to be as low as $n_{at}\approx 10^{12}-10^{13}{\rm cm}^{-3}$ to eliminate the effects of space charge \cite{poprz_pra19} and the macroscopic currents, which can be induced by ionization of atomic outer shells in the environment of the focal spot.}

\begin{figure}[ht!]
\includegraphics[scale = 0.25]{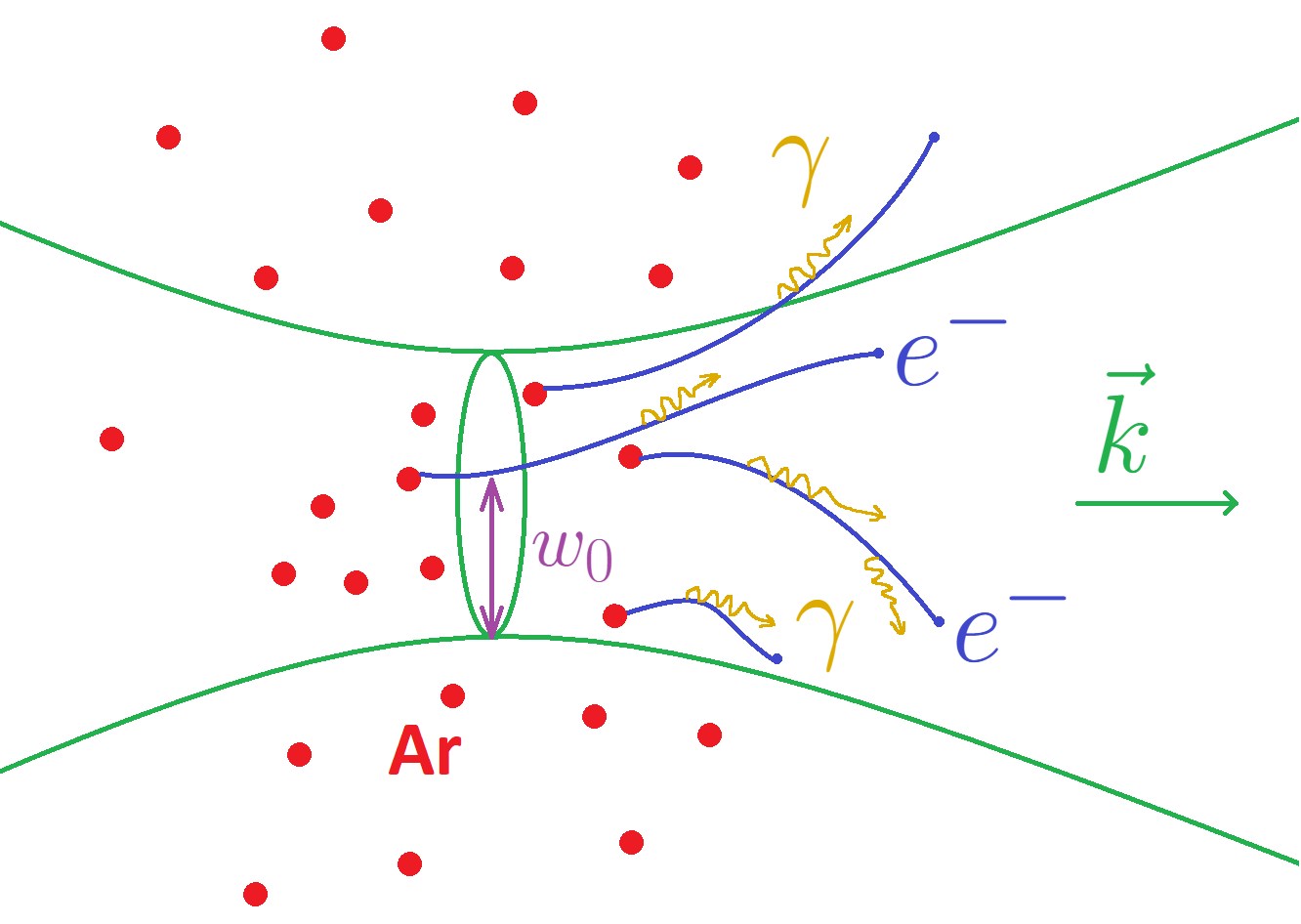}
\caption{Sketch of the interaction setup: a focused laser beam with the wave vector ${\bf k}$; a diluted atomic gas jet crossing the laser focus (red dots); trajectories of electrons released in multiple tunnel ionization of atoms (blue lines); emitted photons (yellow curling lines).}
\label{fig:sketch}
\end{figure}

The gas is ionized by a laser beam focused up to the waist size $w_0\approx 2-4\lambda$ with $\lambda\approx 1\mu{\rm m}$ being the wavelength, which corresponds to the frequency $\omega_L \approx 1.9\cdot 10^{15}$ sec$^{-1}$.
For such moderate focusing, the approximation of a Gaussian beam works acceptably well so that we can use it to estimate the focal Rayleigh length as $z_R=\pi w_0^2/\lambda\approx 30\lambda$ for $w_0=3\lambda$.
This gives for the effective focal volume 
\begin{equation}
    V_{\rm eff}\approx\pi w_0^2 z_R\approx 10^3\lambda^3=10^{-9}{\rm cm}^{-3}
    \label{Veff}
\end{equation}
with $\approx 10^3-10^4$ atoms inside.
For tighter focusing, $w_0\approx\lambda$, the Gaussian form is no longer a good approximation for the description of a stationary laser beam \cite{davis_pra79}. 
In this case one of many known exact solutions (see \cite{scupin_jcp16,narozhny_jetp00} and references therein for examples) can be used to set the field distribution up.

Multi-PW lasers of type \cite{apollon,XCELS,sel_21,ELI-BL} are expected to deliver pulses of $\tau=20-30$fs duration.
 As an example, the current configuration of the Apollon laser facility \cite{apollon_25} delivers 22-fs long pulses at peak power $\approx 2$PW.
In this case, the parameter $kw_0\approx 20$ is smaller than $\omega_L\tau\approx 40-60$, so that solutions found for stationary beams are also approximately valid to describe pulsed fields \cite{narozhny_jetp00}.
Within this approximation, we use the field 
\begin{equation}
\begin{split}
    {\bf E} &=  \frac{{\bf E}_0}{\sqrt{1 + z^2/z_R^2}} 
    \exp\left(-\frac{r^2}{w_0^2\left[1 + z^2/z_R^2\right]}\right) \\
    &\times \cos\left(\omega_L t - kz - \frac{kr^2}{2z\left[1 + z^2/z_R^2\right]} 
    + \arctan\left(\frac{z}{z_R}\right)\right) \\
    &\times f(\omega_L t-kz),\\
    {\bf H} &= -c \cdot \operatorname{rot} \int^t {\bf E}(t^{\prime}) \, dt^{\prime}
\end{split}
\label{gauss-beam}
\end{equation}
of a linearly polarized Gaussian beam with the electric field peak amplitude $E_0$ and a sinusoidal time envelope $f$ of total duration $\tau=2M\pi/\omega_L\equiv MT_L$
\begin{equation}
f(\omega_L t-kz) =
\begin{cases}
    \sin^2\!\left(\frac{\omega_L t-kz}{2M}\right), & \omega_L t-kz \in [0,2M\pi],\\
    0, & \omega_L t-kz \notin [0,2M\pi]~.
\end{cases}
\label{gaussian}
\end{equation}
The integer $M$ gives the total number of optical cycles in the pulse.
In our calculations, we set $M=16$ which gives a full width at half maximum $\tau_{1/2}\approx 25$fs.
In the following, we assume the electric field of the beam (\ref{gauss-beam}) to be polarized along the $x$ axis.
While the electric field is orthogonal to the wave vector ${\bf k}$, the magnetic field defined as (\ref{gauss-beam}) has some small longitudinal component which however makes only a small effect on the dynamics and radiation of photoelectrons.

\subsection{Tunnel ionization}
In this subsection, we introduce the formalism used for description of the ionization step.
In contrast to other sections of this paper where the CGS system is applied, here we use atomic units $e=m=\hbar=1$ with $e$ and $m$ being the elementary charge and electron mass correspondingly.
In these units, the electron charge equals -1, and the speed of light $c=1/\alpha\approx 137$.

We apply the non-relativistic model of tunnel ionization of atomic levels proposed by Smirnov and Chibisov (SC) \cite{smirnov_jetp66} for the static field and by Perelomov, Popov and Terentiev (PPT) \cite{popov_jetp66,perelomov_jetp67,popov_usp04,poprz_jpb14} for the low-frequency field.
The tunnel limit is validated by the smallness of the Keldysh parameter \cite{keldysh_jetp1965, popov_ufn2004}
\begin{equation}
    \gamma_{\rm K}=\frac{\sqrt{2I_p}\omega_L}{E_0}~,
    \label{gamma-K}
\end{equation}
with $I_p$ being the ionization potential.
In infrared laser fields of wavelengths $\lambda\simeq 1\mu{\rm m}$ ionization of atomic levels except for a few outer electrons, always proceeds in the tunnel regime, $\gamma_K\ll 1$. 
For intensities greater than $10^{18}{\rm W/cm}^2$ this condition is well secured with $\gamma_{\rm K}\simeq 10^{-2}-10^{-3}$ or even less \cite{poprz_pra19}.
Relativistic effects on the probability of ionization are negligible up to intensities $\approx 10^{23}{\rm W/cm}^2$ at least \cite{poprz_pra19,poprz-jetpl23}.
Then the magnetic field of the wave does not have an effect on the tunneling process itself, and the rate of ionization is determined by the instant electric field amplitude $E({\bf R}_j,t)$, which also depends on the position of the $j$-th atom ${\bf R}_j$.

\begin{figure*}[ht!]
\includegraphics[width=6.5in]{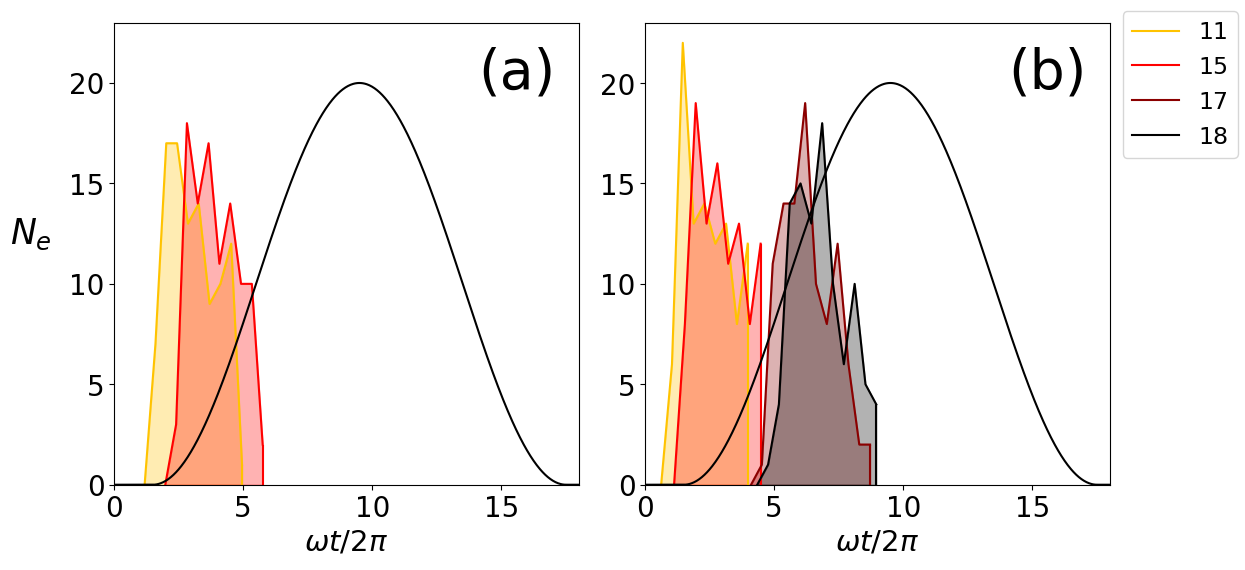}
\caption{Distributions in the number of electrons $N_e$ released through tunneling per interval $\Delta t=T_L/15$ with the rate defined by Eq.(\ref{PPT}) from $N_{\rm at}=100$ ions ${\rm Ar}^{8+}$ randomly distributed in the interaction volume. 
The ion charge is shown by color (see the color coding bars in the right upper corner). The laser field time envelope (\ref{gaussian}) is shown by a black solid line. 
Time is given in laser periods. 
Panel (a) corresponds to intensity ${\cal I}=10^{21}$W/cm$^2$, panel (b) -- to ${\cal I}=10^{22}$W/cm$^2$. 
For the lower intensity, the $1s$ shell is not ionized.}
\label{fig:ionization}
\end{figure*}

We model ionization within the single-electron approximation as a sequential probabilistic process.
 The standard definition of the sequential ionization process assumes that electrons are being removed in the order of increasing ionization potentials.
Recent studies \cite{mironov_pre25} show that this is not always the case; however, deviations from the sequential channel averaged over space and time remain rather minor, and we do not account for them here.
The probability for the outer electron with orbital and magnetic numbers $l$ and $m$ to be removed from its atom (ion) with residual charge $Z_c$ (in these notations, $Z_c=1$ for a neutral atom and ${Z_c}=N$ for $A^{(N-1)+}$ ion) during the time interval $dt$ is given by the SC or PPT static rate \cite{smirnov_jetp66,perelomov_jetp1966,perelomov_jetp67,popov_ufn2004}
\begin{equation}
\begin{gathered}
    dW(t) = w(I_p,\nu,l,m; E(t))dt~, \\
    w = 2^{2\nu+1}C^2_{\nu l} B_{lm} I_p F^{1+|m|-2\nu}(t) \exp\left( -\frac{2}{3F(t)} \right)~.
\end{gathered}
\label{PPT}
\end{equation}
Here
\begin{equation}
  F(t) = \frac{\sqrt{{\bf E}^2(t)}}{(\rm 2 I_p)^{\rm 3/2}}  
  \label{F}
\end{equation}
is the reduced electric field scaled by the characteristic field $E_{\rm ch}=(2I_p)^{3/2}$ of the atomic level (note that the electric field depends also on the atomic position ${\bf R}_j$,
\begin{equation}
    \nu = \frac{Z_c}{\sqrt{2I_p}}
    \label{nu}
\end{equation}
is the effective principal quantum number,
\begin{equation}
    C^2_{\nu l} = \frac{2^{2\nu-2}}{\nu\Gamma(\nu+l+1)\Gamma(\nu-l)}
    \label{C}
\end{equation}
is the squared asymptotic coefficient of the bound state single-electron wave function calculated in Hartree approximation \cite{popov_ufn2004} and the factor
\begin{equation}
    B_{lm} = \frac{(2l+1)(l+|m|)!}{2^{2|m|}|m|!(l-|m|)!}
    \label{B}
\end{equation}
reflects the angular structure of the bound state.
For details of the derivation of the rate formula (\ref{PPT}), as well as for its applications and generalizations we refer the reader to reviews \cite{popov_ufn2004,poprz_jpb14} and references therein.

{ Application of the SC or PPT models for the calculation of ionization events meets two restrictions.
Firstly, at relatively low intensities, when the outer shells are ionized, the sequential channel can be deeply overtaken by the nonsequential one mediated by recollisions \cite{kuchiev_jetpl87,kulander_prl92,corkum_prl93}. 
Because of this, at intensities below $\simeq 10^{17}{\rm W/cm}^2$ and in the field with polarization close to linear the tunnel rates (\ref{PPT}) become inapplicable for the calculation of ion yields.
Although we consider much stronger fields here, moderate intensities, where the recollision mechanism of ionization is relevant, will be reached on the front edge of the pulse or in the pre-pulse, leading to the removal of several electrons with the lowest ionization potentials. 
Secondly, with the field amplitude growing, the regime of ionization can switch from the tunneling to the over-the-barrier mode \cite{mulser_pra99,popov_ufn2004,kostyukov_pra2018}.
Then the SC and PPT formulas again become inapplicable and overestimate the rate by several times for $F=F_*=0.1...0.2$ and higher fields, see examples in \cite{popov_ufn2004}.
The necessity of new formulas \cite{mulser_pra99,tong_jphysb2005,kostyukov_pra2018} extrapolating the tunnel rates to the over-the-barrier domain is determined by the effective value of the reduced field $F_{\rm eff}$ required to deplete the level.
The latter is determined by the level quantum numbers and by the pulse duration, which lowers $F_{\rm eff}$ for smoother time envelopes.
We have checked that (a) for the pulse parameters used in our calculations $F_{\rm eff}\approx 0.05<F_*$, and (b) ionization dynamics for highly charged states with $Z_c\ge 9$ is weakly sensitive to the initial charge state.
This makes application of the tunnel rates well justified.
}

{ Taking into account this argumentation, we perform calculations for ions Ar$^{8+}$ with 10 electrons remaining in the shell and the ionization potentials varying from $I_p(Z_c=9)= 422.5$ eV for the outermost electron to $I_p(Z_c=18)=4426$ eV of the H-like ion.} 
Atoms were randomly distributed in a cubic volume of size $w_0$ in each direction  centered in the focal point; calculations were performed from the time instant $t_1=-3T_L/2$ when the front edge of the ionizing pulse arrives in the plane ($z = -w_0/2$) to $t_2 = 50T_L$, where $T_L = 2\pi/\omega_L$ is the laser period. 
The time step $dt = T_L/40$. 
At each time interval, an ionization event may occur according to the random number generator applied for each atom with the probability density given by \eqref{PPT}.

{ Fig.2 shows the temporal ionization dynamics of a pre-ionized ${\rm Ar}^{8+}$ target.
The distributions were obtained for an ensemble of 100 atoms randomly placed in the central part of the focus, which corresponds to a concentration $n_{\rm at}\approx 4\cdot 10^{12}{\rm cm}^{-3}$.
The residual ion charge is color-coded.
The distributions are weakly sensitive to the initial ionic state, thus we expect them to be similar to the case of neutral argon, although the rate formula \eqref{PPT} does not correctly describe ionization of the outer orbitals.
As we will see in Section III below, the dominant contribution to emission of high-energy photons comes from the $1s$ electrons.
They are not present in Fig.2(a), because of an insufficient intensity.
The plots of Fig.2(b) show that these electrons are removed at ${\cal I}\approx 2.5\cdot 10^{21}{\rm W/cm}^2$ (cf. Fig.4 in \cite{poprz_pra19}).
This gives an estimate for the effective reduced field $F_{\rm eff}\approx 0.05$, which justifies the use of the semiclassical tunnel rate (\ref{PPT}).
}

\subsection{Post-ionization photoelectron dynamics}

After the ionization event, an electron moves along a classical trajectory determined by the Lorentz force
\begin{equation}
    \frac{d{\bf p}}{dt}=e\bf{E} + \it\frac{e}{c} \bf{v}\times{\bf H}
    \label{eqm}
\end{equation}
with initial conditions ${\bf r}(t_{jn})={\bf R}_j$, ${\bf v}(t_{jn})=0$.
Here the subscribe $j$ numbers the atom, and $n$ – the electron in the $j$-th atom, $n=9,...,18$ for the considered case.
On the femtosecond time scale, we discard the atomic motion inside the focal spot. 
The zero initial velocity condition stems from the model of nonrelativistic tunneling in a quasi-static field introduced in the previous subsection.
In combination with the high field intensity, this initial condition results in a specific form of trajectories, which are strongly elongated in the direction of the laser pulse propagation and experience only a few oscillations before leaving the focus.
Several trajectories illustrating this feature are shown in Fig.3.
Their color code, identical to that of Fig.2, corresponds to the charge state and helps to approximately identify the instant of ionization.
While the ``early'' trajectories corresponding to electrons with low ionization potentials experience one or two oscillations before their escape from the focus, for the ``late'' trajectories no single oscillation occurs.
Note that the photoelectron trajectories have a non-vanishing $y$-component, despite the zero initial velocity and linear polarization of the electric field along the $x$-axis.
This is due to the ponderomotive scattering in the focused pulse, where the magnetic field possesses a small longitudinal component. 

\begin{figure*}[ht!]
\includegraphics[ width=6.5in]{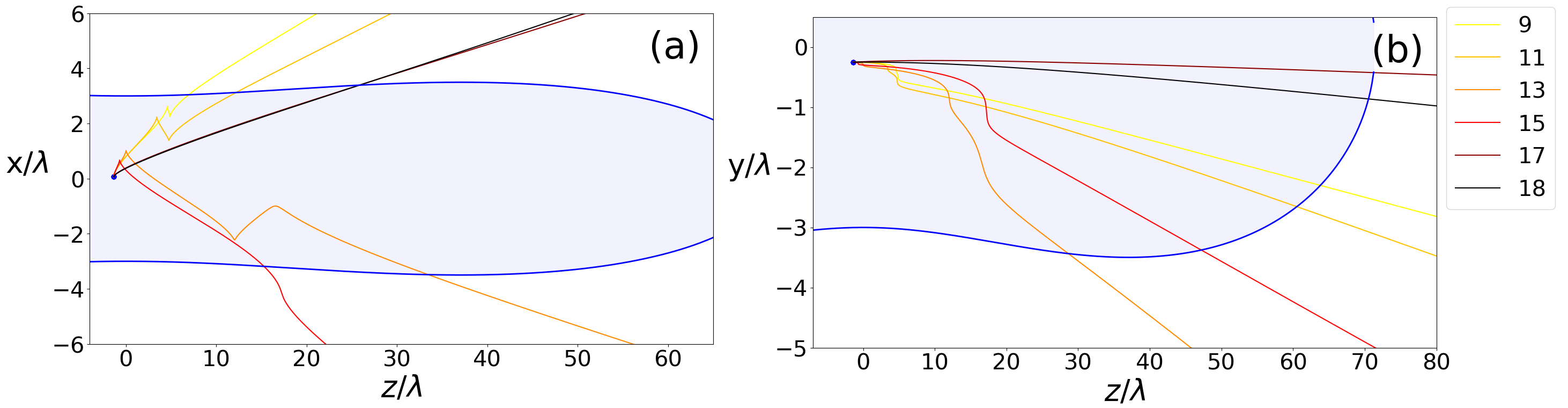}
\caption{Electron trajectories in the $(x,z)$ (a) and $(y,z)$ (b) planes for an atom with coordinates $x_0=0.06\lambda,~y_0=-0.25\lambda,~z_0=-1.35\lambda$. The central part of the focus with ${\cal I}\ge {\cal I}_0/{\rm e}^2$ is shown blue. 
The color code for the electron number inside the shell is shown in the upper right corner.}
\label{fig:trajectories}
\end{figure*}

\begin{figure*}[ht!]
\includegraphics[width=7.3in]{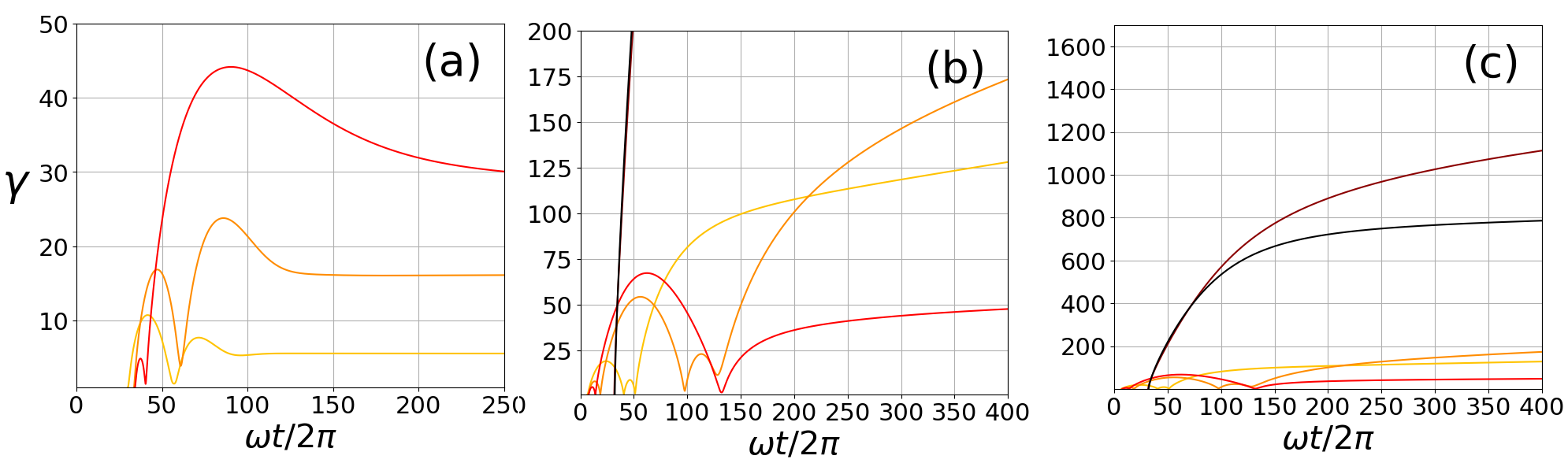}
\caption{Time-dependent Lorentz factors $\gamma(t)$ for 10 electron trajectories with the parameters of Fig.2. Panels (a) and (b) correspond to those of Fig.2. Panel (c) shows the same as (b) in a different scale to make the trajectories of the two $1s$ electrons visible.
The electron number inside the shell is coded by the same colors as in Fig.2.}
\label{fig:gamma}
\end{figure*}

For a Gaussian-like beam with the waist radius $w_0 = N\lambda$ we consider an electron to have escaped from the focal area when any of the two conditions
\begin{equation}
    x \geqslant N\lambda~,~~~z \geqslant z_R = \frac{\pi w_0^2}{\lambda}=\pi N^2\lambda~.
    \label{xz}
\end{equation}
is fulfilled.
To estimate the time spent by an electron inside the focus, and the respective phase difference between the instant of ionization and that of the exit, we describe the motion using the plane wave approximation.
Assuming for the vector potential to have the form
\begin{equation}
    A_x = \frac{cE_0}{\omega}g(\varphi)~,~~~\varphi = \omega_L t - kz
    \label{Ax}
\end{equation}
with $g=f\cos\varphi$, and $f$ given by (\ref{gaussian}) or any other pulsed function, one obtains the momenta and coordinates
\begin{equation}
    \frac{p_x}{c} = a_0(g(\varphi_0) - g(\varphi))~,~~~\frac{p_z}{c} = \frac{a_0^2}{2}(g(\varphi_0) - g(\varphi))^2
    \label{pxpz}
\end{equation}
\begin{equation}
    x - X_j\approx a_0\lambda\frac{g'(\varphi_0)}{2}(\varphi - \varphi_0)^2~,~~~z-Z_j\approx a_0^2\lambda\frac{[g'(\varphi_0)]^2}{6}(\varphi - \varphi_0)^3
    \label{xz}
\end{equation}
Expressions (\ref{xz}) are approximate and valid under the condition $\varphi-\varphi_0\ll 1$, which is justified by the form of trajectories shown in Fig.3.
In Eqs.(\ref{pxpz}) and (\ref{xz}) the invariant dimensionless field amplitude is introduced 
\begin{equation}
    a_0 = \frac{eE_0}{mc\omega_L}~.
    \label{a0}
\end{equation}
Then the exit condition is either $x(\varphi_x)\simeq N\lambda$ or $z(\varphi_z)\simeq z_R$.
This gives for the respective phase intervals
\cite{kalymbetov_qe21}
\begin{equation}
    \Delta\varphi_x\equiv \varphi_x  - \varphi_0\simeq \sqrt{\frac{4\pi N}{a_0}}~,
    \label{dphix}
\end{equation}
\begin{equation}
    \Delta\varphi_z\equiv \varphi_z  - \varphi_0\simeq \left(\frac{12\pi^2N^2}{a_0^2}\right)^{1/3}~.
    \label{dphiz}
\end{equation}
Both values are small at $a_0\gg 1$, which explains the trajectory shape with less than one oscillation.
With the field intensity growing, the electrons tend to escape from the front edge of the laser focus, as the interval (\ref{dphiz}) vanishes faster.
Equalizing (\ref{dphix}) and (\ref{dphiz}) one obtains an estimate of the field amplitude $a_0\simeq 20$ above which the longitudinal escape focus becomes dominant.
In agreement with this qualitative estimate, the trajectories shown in Fig.3 escape rather  longitudinally than laterally for $1s$ electrons released from their ions near the field maximum.  
The angle between the propagation direction and the electron velocity upon escape from the focus can be estimated as 
\begin{equation}
    \theta_0\approx\frac{p_x(\Delta\varphi_z)}{p_z(\Delta\varphi_z)}\simeq\frac{0.4}{N^{2/3}a_0^{1/3}}~.
    \label{theta}
\end{equation}
Note that for a linearly polarized field of wavelength $\lambda=1\mu{\rm m}$ and $a_0=100$ corresponds to the intensity ${\cal I}\approx 2\cdot 10^{22}$W/cm$^2$.

\subsection{Radiation}

For each trajectory ${\bf r}_{jn}(t)$, a spectral-angular distribution of the emitted radiation can be calculated numerically using the well-known expression \cite{Jackson,Landau2}
\begin{equation}
    \frac{d{\cal{E}}_{jn}}{d\omega dO_{\bf n}}=   \frac{e^2\omega^2}{4\pi^2c^3} \Bigg|\int_{-\infty}^{\infty} dt [{\bf n} \times [{\bf n} \times {\bf v}(t)]] e^{i\omega[t-{\bf n}\cdot{\bf r}_{jn}(t)/c]} \Bigg|^{\rm 2}~.
    \label{sad}
\end{equation}
Here $dO_{\bf n}$ is the solid angle in the direction defined by the unit vector ${\bf n}$.
{The electron velocity ${\bf v}=0$ for $t\le t_{jn}$, where $t_{jn}$ is the corresponding ionization instant.}
Due to the high electron $\gamma$-factors we assume radiation to be completely incoherent so that the integral signal of the atomic ensemble is obtained by summing (\ref{sad}) over all individual contributions.

In addition to the radiation in the field of the Gaussian pulse (\ref{gauss-beam}), we will consider that in the field of a counter-propagating weakly focused pulse.
Its field can be treated as a plane monochromatic wave, and the spectral-angular distributions can be calculated with the help of analytic expressions for differential cross-sections of nonlinear Thomson scattering derived in the pivotal paper by Sarachik and Schappert \cite{sarachik_prd70}.
To this end, we will use the Lorentz transformation to a reference frame (R-frame) where an individual electron trajectory has zero average velocity.
In this R-frame, radiation consists of harmonics of the fundamental frequency  
\begin{equation}
    \omega_R=\frac{\omega_{L}'-{\bf V}_0{\bf k}}{\sqrt{1 - V_0^2/c^2}}~,
    \label{omR}
\end{equation}
where $\omega_{L}'$ is the frequency of the counter-propagating beam in the laboratory frame (L-frame) and ${\bf V}_0$ is the relative velocity of R-L, whose value is the asymptotic velocity of the electron after it leaves the field of the ionizing pulse.
In the R-frame, the angular distribution of the power emitted at $\omega_n=n\omega_R$ is given by \cite{sarachik_prd70} 
\begin{equation}
    \begin{gathered}
        \frac{d{\cal{P}}^{(n)}}{dO}(\vartheta,\phi) = \frac{(e\omega a_0 n)^2}{8\pi c (1 + a_0^2)} \times \\
        \times ( |\chi_2^{n}|^2 + \frac{1}{4}a^2|\chi_3^{n}|^2 - |\chi_2^n\cos\phi\sin\vartheta - \frac{1}{2}a\chi_3^n\cos\vartheta|^2)
    \end{gathered}
\label{R-distr}
\end{equation}
with
\begin{equation}
\begin{gathered}
    \chi_2^{n} = \sum_{l=-\infty}^{\infty} J_l\left[na^2\sin\frac{\vartheta^2}{2}\right] \times \\
    \quad \times \left((-i)^{n+2l+2}J_{n+2l+1}(\rho) + (-i)^{n+2l-2}J_{n+2l-1}(\rho)\right), \\
    \chi_3^{n} = \sum_{l=-\infty}^{\infty} J_l\left[na^2\sin\frac{\vartheta^2}{2}\right] \times \\
    \quad \times \left((-i)^{n+2l+2}J_{n+2l+2}(\rho) + (-i)^{n+2l-2}J_{n+2l-2}(\rho)\right), \\
    a^2 = \frac{a_0^2}{4 + 2a_0^2}, \quad \rho = 2an\sin\vartheta\cos\phi~.
\end{gathered}
\label{chi-R}
\end{equation}
Here $a_0$ is the dimensionless invariant field amplitude (\ref{a0}), $\vartheta$ and $\phi$ are the polar and azimuthal angles with respect to the x-axis pointing along the electric field $\bf E$ polarization direction in R-frame.

In order to calculate the distribution in the direction ($\vartheta$,$\phi$) in the L-frame, we perform the following steps: (a) rotate the coordinate system ($\vartheta$,$\phi$) $\rightarrow$ ($\vartheta'$,$\phi'$) to direct the $z'$ axis along the electron velocity $\bf V_0$ in the L-frame; (b) apply the Lorentz transform from the L- to R-frame:
\begin{gather}
    \phi'_R = \phi' \\
    \nonumber
    \cos\vartheta'_R = \frac{\cos\vartheta' - V_0/c}{1 -  V_0/c\cos\vartheta'};
    \label{theta-R}
\end{gather}
(c) rotate the coordinate system ($\vartheta'_R$,$\phi'_R$) $\rightarrow$ ($\vartheta''_R$,$\phi''_R$) so that the choice of axes corresponds to that assumed in Eq.(\ref{R-distr}); (d) perform the inverse Lorentz transformation to the L-frame. 
This procedure is described in details in \cite{sarachik_prd70} (see Eqs. 3.19 - 3.32b there). 
As a result, the angular distribution of the radiated power in the L-frame and the spectrum have the form:
\begin{equation}
    \frac{d{\cal{P}}}{dO}(\vartheta,\phi) = \frac{(1 -\frac{V_0^2}{c^2} )^2}{(1 - \frac{V_0}{c}\cos{\vartheta'})^3}\sum_{n = 1}^{\infty}\frac{d{\cal{P}}^{(n)}}{dO}\left(\vartheta''_R(\vartheta,\phi),\phi''_R(\vartheta,\phi)\right)
    \label{P-n}
\end{equation}
\begin{equation}
    \omega_n = n\omega_R\frac{\sqrt{1-\frac{V_0^2}{c^2}}}{1 - \frac{V_0}{c}\cos\vartheta'}.
    \label{P-om}
\end{equation}
Eqs.(\ref{P-om}) assume a monochromatic probe field of infinite time duration.
Thus, for our problem, these expressions are approximate and discard the line broadening due to the finite spectral width of the probe pulse. 
To estimate the angular distribution of the emitted energy we multiply the power distribution (\ref{P-n}) by the counter-propagating pulse duration $\tau^{\prime}$. \\

\section{Results and discussion}
\subsection{Radiation in the field of the ionizing pulse}
When only the ionizing laser pulse is considered, we calculate the angular-frequency distribution of the radiated energy numerically, along Eq.(\ref{sad}).
Quite expected, the results show that the main contribution into the radiation signal is made by the two 1s electrons detached at ${\cal I}\approx (2.5-4.0)\cdot 10^{21}{\rm W/cm}^2$, before the field maximum (see Fig.2(b)).
{ Ionization potentials of the 1s shell are significantly higher than those of the other sub-shells, $I_p(1s^2)=4121$eV and $I_p(1s^1)=4426$eV, correspondingly, so that these electrons can survive on their orbitals almost till the field maximum.
As a result, the post-ionization motion is highly co-propagating and in-phase (see the black curves in Fig. \ref{fig:trajectories}), and the electrons reach $\gamma\approx 10^3$. 
In contrast, for ${\cal I}=10^{21}$W/cm$^2$ ionization of the $1s$ shell does not occur at all, see Fig.2(a), and the photoelectrons escape the focus mostly laterally having $\gamma\approx 30$.}
The resulting angular distribution of emitted photons is mostly concentrated in the $(x,z)$ plane and narrow-peaked around an angle $\theta_0$, which can be estimated from (\ref{theta}).
{ This estimation is sufficiently accurate for $a_0 \sim 10^2$, i. e., in the interval where ionization of the 1s shell already occurs near the field maximum.}

\begin{figure*}[ht!]
\includegraphics[ scale = 0.4]{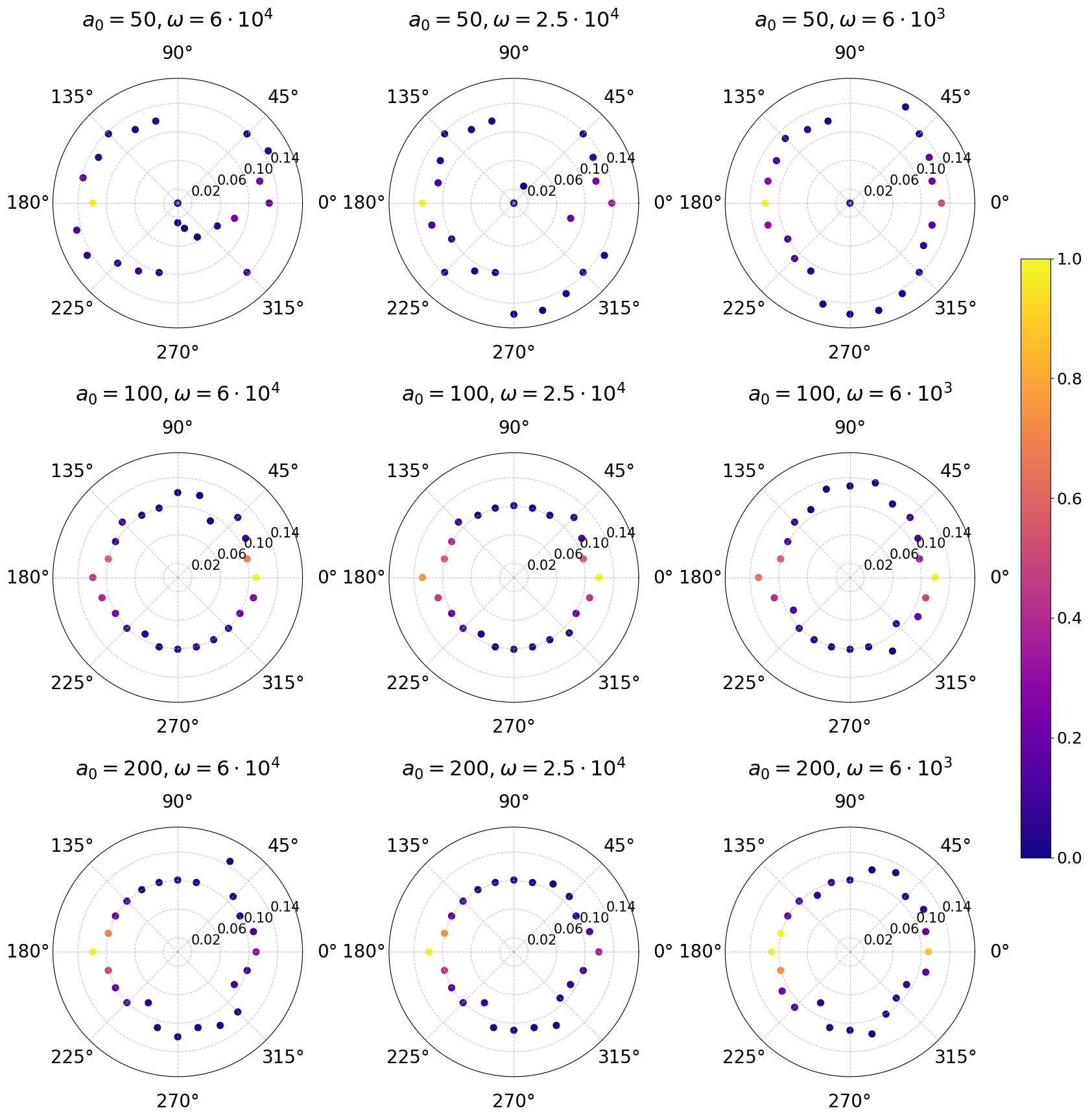}
\caption{ { Dependence of the polar angle $\theta_0$ between the laser wave vector $\bf k$ and the maximum of the angular distribution at fixed frequency $\omega$ on the azimuthal angle in the plane orthogonal to $\bf k$ for $a_0=50$ (upper row), $a_0=100$ (middle row) and $a_0=200$ (lower row). The radiation frequency is $6\cdot10^4\omega_L$ (left column), $2.5\cdot 10^4\omega_L$ (middle column) and $6\cdot10^3\omega_L$ (right column). The electric field polarization direction is horizontal $(\psi=0,\pi)$. The relative radiation power is color-coded.}}
\label{fig:theta0_dependence}
\end{figure*}

It is instructive to look at the angular distribution in the plane perpendicular to the propagation direction for a fixed photon energy.
Fig.\ref{fig:theta0_dependence} shows the dependence of the value of the polar angle $\theta_0$ where the distribution reaches its maximum on the azimuthal angle $\psi$ for $a_0 = 50, 100$ and 200 at different values of the photon frequency.
The electric field polarization is horizontal and corresponds to $\psi=0,\pi$.
The absolute magnitude of the distribution is color-coded.
{Here we could make the following observations.
\begin{enumerate}
    \item {The distribution for $a_0=50$ is much less regular than for higher field amplitudes, particularly at $\omega=6\cdot 10^4\omega_L$, because for this lowest intensity the selected frequencies correspond to the high-energy tail of the spectrum, see Fig.6 below.}
    \item {With $a_0$ growing and for the correspondingly increasing frequencies, the value of $\theta_0$ follows approximately the estimate (\ref{theta}). This tendency is best seen for the distributions placed on the anti-diagonal of the plot.}
    \item{The maximum is achieved in the polarization direction, although the distribution in $\psi$ appears rather broad.}
\end{enumerate}
}

\begin{figure*}[ht!]
\includegraphics[ scale = 0.37]{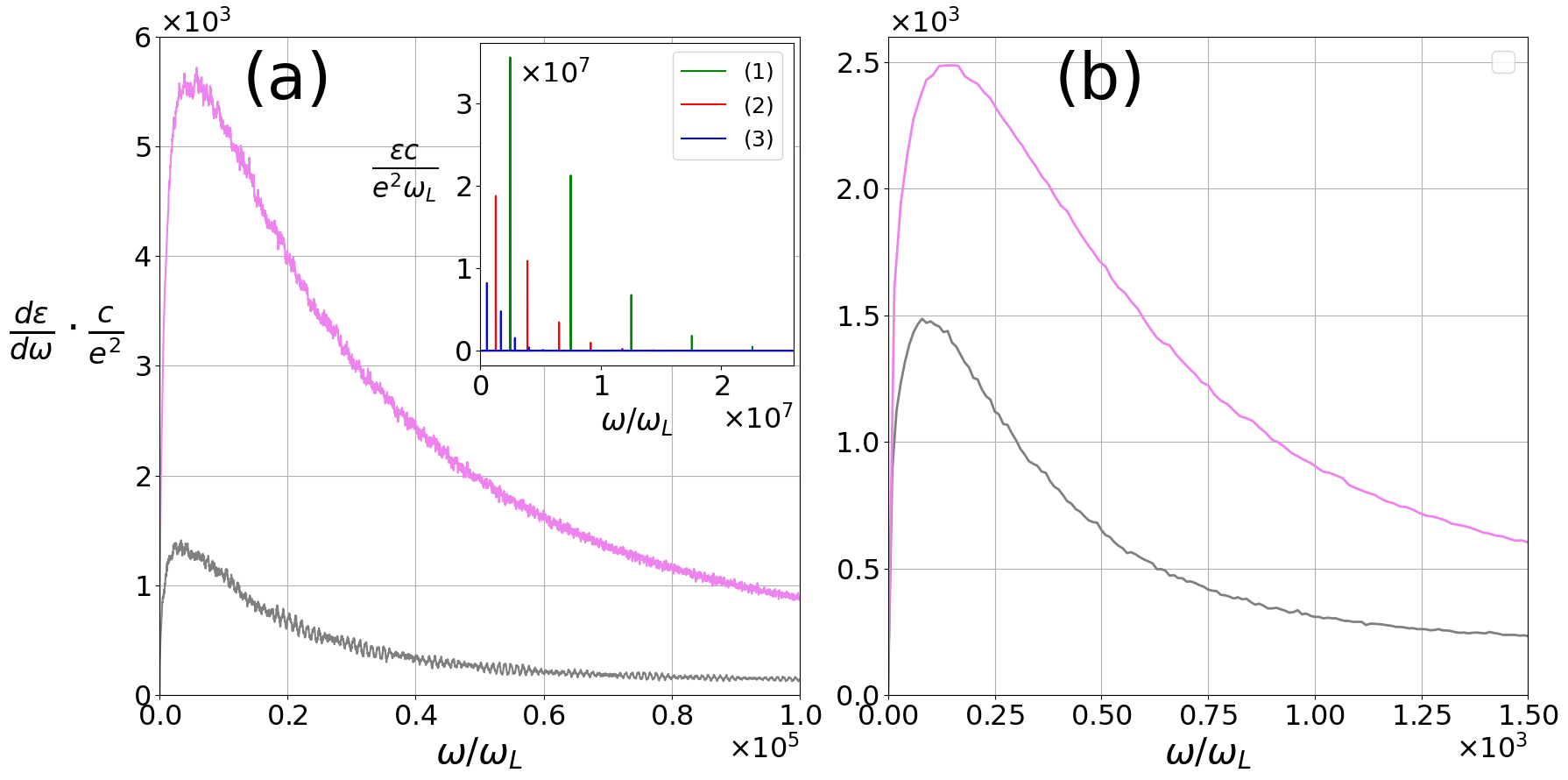}
\caption{Angle-integrated spectra of radiation emitted by an ensemble of 100 atoms at $a_0 = 100$ (pink curves), $a_0 = 50$ (gray curves). { Panel (a) shows the spectra radiated solely by $1s$ electrons, while (b) shows the contribution of all other electrons.} The spectra are scaled by  $e^2/c=7.68\cdot 10^{-30}$erg$\cdot$c. 
Harmonics emitted by three electrons in the field of the counter-propagating pulse with $a_0^{\prime} = 1$, same frequency $\omega_L^{\prime}=\omega_L$ and duration $\tau' = 30$fs are shown in the insert. In this case, the spectrum is scaled by $e^2\omega_L/c=1.53\cdot 10^{-14}$erg. The $\gamma$-factors of the three electrons after their escape from the ionizing pulse are: (1) $\gamma = 796$ (2) $\gamma = 574$, (3) $\gamma = 381$.}
\label{fig:spectral}
\end{figure*}

The angle-integrated spectral distribution of radiation is shown in Fig. \ref{fig:spectral} for $a_0=50$ and $a_0=100$. {In order to analyze the role of inner orbitals with high ionization potentials, we plotted separately the contribution of the $1s$ electrons (panel (a)) and that of all other electrons (panel (b)).
The difference between the two sets of plots is noticeable.
While the contribution of the outer shells into the radiation power grows by factor $\approx 1.7$ with increasing of $a_0$ from 50 to 100, that of the $1s$ shell increases by factor $\approx 4$ making it dominant at $a_0=100$.
This fast growth is caused by a rapid increase in the number of fully ionized atoms, which is a highly nonlinear function of the field amplitude \eqref{PPT}.
}

For $a_0=100$, the total emitted energy per atom is:
\begin{gather}
    {\cal{E}} = \frac{1}{N_{\text{at}}}\int_0^\infty \frac{d{\cal{E}}}{d\omega}d\omega \approx 2\cdot 10^4 \ \text{eV}.
    \label{rad}
\end{gather}
Then the number of photons emitted per atom can be estimated as
\begin{gather}
    N_{\gamma} = {\cal{E}} / \hbar\omega_{\rm ch} \simeq 2-3,
    \label{ngamma}
\end{gather}
where the characteristic frequency $\omega_{\rm ch}\approx 0.5\cdot 10^4\omega_L$ corresponds to the maximum of the purple curve in Fig.6(a).
{ Note also the different horizontal scales in panels (a) and (b), meaning that radiation with $\omega\simeq 10^4\omega_L$ is predominantly emitted by $1s$ electrons.}

Summarizing this subsection, the angular distributions of radiation resulting from tunneling ionization are sensitive to the laser field amplitude at the space-time point, where ionization occurs. 
This dependence is nonlinear and therefore survives the averaging over the focal volume.
The contribution of inner shells into the total emitted energy is pivotal and grows quickly when the field intensity passes the corresponding ionization threshold. 
Observation of these features of radiation can serve as an additional tool for the {\it in situ} measurement of the peak intensity of multi-PW laser beams. 
However, estimations \eqref{rad}, \eqref{ngamma} show that the emitted energy and the number of  photons per atom are rather low. 
Both of these values are significantly suppressed as a result of the co-propagating motion of electrons. 
Therefore, it is of interest to investigate how radiation can be enhanced by a collision with a relatively weak counter-propagating pulse. 

\subsection{Radiation in a weak probe pulse}

\begin{figure}[ht!]
\includegraphics[ width=3.5in]{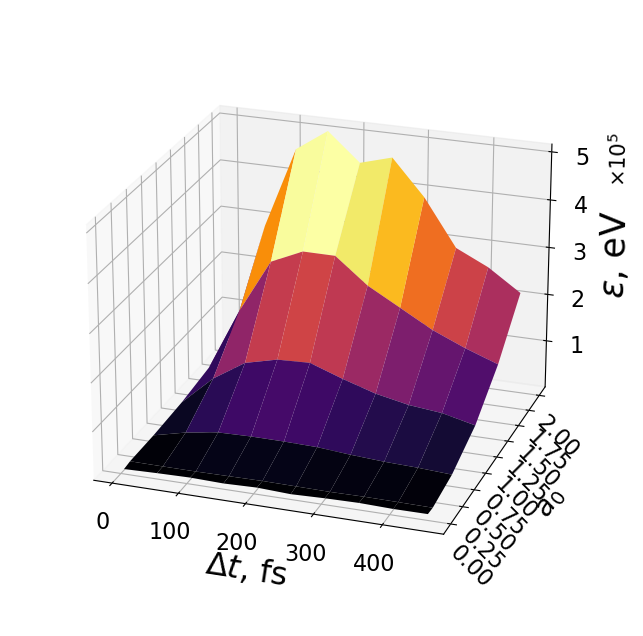}
\caption{Energy emitted per atom as a function of the time delay $\Delta t$ between the maxima of the ionizing and the probe pulses at the focal plane and on the probe pulse amplitude. The ionizing pulse waist radius $w_0 = 3 \lambda$, that of probe pulse $w_0^{\prime} = 9 \lambda$.}
\label{fig:time_delay}
\end{figure}

To enhance the signal, we consider the interaction of the longitudinally accelerated photoelectrons with a counter-propagating weakly focused pulse (probe pulse). 
As we show below, the amplitude $a_0 \simeq 1 $ is sufficient for a significant enhancement. 
Assuming an experiment, where a multi-PW pulse is split into two parts with the energy ratio 1:100  propagating along two beamlines, one may consider the probe focal spot of roughly 10 times larger area than that of the ionizing pulse (i.e., the focal waist of the probe pulse $w_0^{\prime}\approx 3w_0$).
In order to find optimal conditions for radiation of the accelerated photoelectrons we introduce a variable time delay $\Delta t$ between the instants when the front edges of the ionizing and the probe pulses arrive at the focal plane.

{ In contrast to the calculations presented above, where the spatial field distribution of the ionizing pulse played a significant role in dynamics and radiation of the photoelectrons, here owing to the weak focusing of the probe pulse we can use the plane wave approximation and the respective Eqs.(\ref{omR})-(\ref{P-om}) to compute the radiation spectrum.}
The insert in Fig.6 shows the angle-integrated spectra calculated analytically.
In the forward direction, radiation is emitted  at frequencies $\omega_{\rm ch}\simeq 10^6\omega_L$. The total radiated energy per atom in the field of the probe pulse is
\begin{gather}
    {\cal{E}} = \frac{1}{N_{\text{at}}} \sum_{n = 0}^{\infty} \int dO \frac{d\mathcal{E}_n}{dO} \simeq 10^5 \text{eV},
    \label{rad_co}
\end{gather}
where the emitted energy ${\cal{E}}_n$ is estimated by multiplying the power (\ref{P-n}) by the probe pulse duration $\tau' = 30$fs.
{ Note that backscattering in the field of a plane wave occurs at odd harmonics \cite{sarachik_prd70}.
Because the photoelectrons propagate, after leaving the ionizing field, almost along the pulse axis, see Eq.(\ref{theta}) and Fig. \ref{fig:trajectories}, the peaks visible in the insert of Fig.6 are separated by the doubled frequency of the fundamental harmonic (that of the first peak). 
Even harmonics are also present, although with much lower intensities.
As an example, the intensity of the second harmonic is two orders of magnitude lower than that of the first.}

Eqs.\eqref{R-distr}-(\ref{P-om}) do not take into account the finite size of the pulse in the longitudinal direction so that (\ref{rad_co}) gives only an estimate of the full emitted energy. 
When the total emitted energy is only concerned, it can be easily found by integration of the radiation power along the trajectory \cite{Landau2}
\begin{equation}
    \mathcal{E}= \frac{2e^4}{3m^2c^3}\int\limits_{-\infty}^{+\infty} \frac{\left\{{\bf{E}} + {\bf{v}}\times{\bf{H}}/c\right\}^2 - ({\bf{Ev}})^2/c^2}{1-v^2/c^2}dt.
    \label{P_inField}
\end{equation}
Here ${\bf v}(t)=0$ before the ionization instant, as in (\ref{sad}).

Fig.7 shows the energy emitted per atom as a function of the time delay between the maxima of the pulses at the focal plane and of the amplitude of the probe pulse $a_0^{\prime}=0.0-2.0$. 
The amplitude of the ionizing pulse is $a_0=100$.
The optimal time delay $\Delta t\approx 200$fs corresponds to the interaction scenario, when the photoelectrons reach their maximal $\gamma$-factors right before the interaction with the probe pulse begins and before their lateral coordinates exceed the probe pulse waist $w_0^{\prime}=9\lambda$.

\section{Conclusions and outlook}
We considered radiation, which accompanies tunnel ionization of heavy atoms in focused laser pulses of intensity $10^{21}{\rm W/cm}^2$ and higher. 
Although this radiation is described within the standard framework of classical electrodynamics, it demonstrates specific features different from those typical for radiation of an incident electron beam colliding with a strong electromagnetic pulse.
This difference mostly stems from the ionization initial condition, which assumes zero or small electron velocity upon its liberation from the parent ion inside the focus.
The very anisotropic angular distribution is highly peaked along the slant line of a cone whose angle depends on the intensity of the ionizing pulse in the central part of the focus.
{ The main contribution to the radiation spectrum comes from ionization of $1s$ orbitals, which occurs when the laser intensity exceeds the corresponding threshold. 
This makes both the magnitude and the width of these spectra highly sensitive to the peak value of intensity in the central part of the laser focus.}
This sensitivity can be used for diagnostics of the laser intensity in the focus of a high-power laser beam, complimentary to the ionization diagnostic discussed earlier \cite{walker_pra01,link_instr06,poprz_pra19,ciappina_lpl2020}.

Because of the co-propagating electron trajectories, the emitted energy appears low even for intensities $\simeq 10^{22}{\rm W/cm}^2$, but can be considerably enhanced by a relatively weak counter-propagating loosely focused pulse, which can be prepared by splitting the initial multi-PW beam in two coherent sub-beams of different energy.
{ According to (\ref{Veff}), the total number of atoms in the interaction volume is $\simeq 10^3-10^4$, which gives the total number of high-frequency photons emitted per laser shot to be of the same order of 2-3 times higher (\ref{ngamma}).
Modern x-ray detectors (see, e.g. \cite{detectors_sr02,detectors_fp23}) are capable of registering such low photon fluxes in a wide spectral range, which covers that $\hbar\omega=10^3-10^6$eV emerged in our calculations, and with high angular resolution.}

Except for the development of {\it in situ} diagnostics of multi-PW or exawatt laser radiation, the presented results can be of use for the further elaboration of the theory of quantum-electrodynamic cascades initiated by ionization in laser fields of extreme intensity \cite{artemenko_pra2017}.

\section*{Acknowledgment}
Authors are thankful to A.M. Fedotov for valuable discussions and advice. 
NVM acknowledges financial support from the Basis foundation in the part of work related to numerical simulations and analysis of the obtained results.
SVP acknowledges financial support from the Russian Science Foundation, grant No.25-12-00336, in the part of work that includes formulation of the problem, qualitative analysis of equations, and preparation of the manuscript.

\bibliography{bibliography}

\end{document}